\documentclass[prl,twocolumn,a4paper,showpacs]{revtex4}
\usepackage{graphicx}
\begin{document}

\title{`Algorithmic cooling' in a momentum state quantum computer}
\author{Tim Freegarde}
\email{tim.freegarde@physics.org}
\affiliation{Dipartimento di Fisica, Universit\`{a} di Trento, 38050 Povo (TN), Italy}
\author{Danny Segal}
\affiliation{Quantum Optics and Laser Science, Imperial College, London SW7 2BZ, U.K.}

\date{\today}

\begin{abstract}
We describe a quantum computer based upon the coherent manipulation
of two-level atoms between discrete one-dimensional momentum states.
Combinations of short laser pulses with kinetic energy dependent free
phase evolution can perform the logical invert, exchange, CNOT and
Hadamard operations on any qubits in the binary representation of the
momentum state, as well as conditional phase inversion. These allow a
binary right-rotation, which halves the momentum distribution in a single
coherent process. Fields for the coherent control of atomic momenta
may thus be designed as quantum algorithms.
\end{abstract}

\maketitle

Proposed schemes for quantum computation~\cite{Deutsch85} have tended,
quite naturally, to focus on quantum analogues of classical binary computing
elements. The nuclear spins of a molecule, or of an ensemble of trapped
atoms or ions, thus mimic the bits of a conventional computer. In this article,
we address a less obvious system, in which information is represented by
the momentum of a single atom or molecule, which is manipulated using
laser pulses in a one-dimensional geometry that restricts each species to a
ladder of equally spaced momentum states. Although any one laser pulse can
change the species momentum, through photon absorption and stimulated
emission, by only a single photon impulse, we find that sequences of
pulses, interspersed with periods of momentum-dependent phase evolution,
allow a full suite of quantum computational operations on the qubits
comprising the binary representation of the momentum state.

The size of the momentum state quantum computer grows in proportion to the
number of quantum states included, where conventional candidates instead
scale with the number of qubits representing those states. The number of laser
pulses needed to perform each logical operation increases similarly and,
although the overall duration proves to be less drastically affected, momentum
state systems therefore hold limited promise for real computing. Nonetheless,
the scheme outlined here is based upon simple and readily-available elements,
albeit in complex combinations, and could thus complement NMR
systems~\cite{Chuang98,Jones98b} as a testbed for experimental studies.

It is, however, in the design of complex fields for coherent control that we
forsee the greatest potential, for if momentum-changing operations can
form the basis of a quantum computer then the pulse sequences for
optical manipulation may be optimized as quantum computational
algorithms. In this respect, the momentum state quantum computer is an
enthusiastic development of schemes for interferometric
cooling~\cite{Weitz00} and the coherent amplification of laser
cooling~\cite{Freegarde02c}.

\begin{figure}
\setlength{\unitlength}{1mm}
\begin{picture}(86,36)
{\thicklines \multiput(47,0)(0,5){8}{\line(1,0){5}} }
\put(39,0){ \makebox[5mm][r]{$-2\hbar k$} }
\put(39,5){ \makebox[5mm][r]{$-\hbar k$} }
\put(39,10){ \makebox[5mm][r]{$0$} }
\put(39,15){ \makebox[5mm][r]{$\hbar k$} }
\put(39,20){ \makebox[5mm][r]{$2\hbar k$} }
\put(39,25){ \makebox[5mm][r]{$3\hbar k$} }
\put(39,30){ \makebox[5mm][r]{$4\hbar k$} }
\put(53,0){ \parbox[b]{5mm}{$|g\rangle$} }
\put(53,5){ \parbox[b]{5mm}{$|e\rangle$} }
\put(53,10){ \parbox[b]{5mm}{$|g\rangle$} }
\put(53,15){ \parbox[b]{5mm}{$|e\rangle$} }
\put(53,20){ \parbox[b]{5mm}{$|g\rangle$} }
\put(53,25){ \parbox[b]{5mm}{$|e\rangle$} }
\put(53,30){ \parbox[b]{5mm}{$|g\rangle$} }
\put(29.5,0){\makebox[2mm][c]{$\vdots$}}
\put(25,5){\makebox[2mm][c]{$1$}} \put(28,5){\makebox[2mm][c]{$1$}} \put(31,5){\makebox[2mm][c]{$1$}} \put(34,5){\makebox[2mm][c]{$1$}}
\put(25,10){\makebox[2mm][c]{$0$}} \put(28,10){\makebox[2mm][c]{$0$}} \put(31,10){\makebox[2mm][c]{$0$}} \put(34,10){\makebox[2mm][c]{$0$}}
\put(25,15){\makebox[2mm][c]{$0$}} \put(28,15){\makebox[2mm][c]{$0$}} \put(31,15){\makebox[2mm][c]{$0$}} \put(34,15){\makebox[2mm][c]{$1$}}
\put(25,20){\makebox[2mm][c]{$0$}} \put(28,20){\makebox[2mm][c]{$0$}} \put(31,20){\makebox[2mm][c]{$1$}} \put(34,20){\makebox[2mm][c]{$0$}}
\put(25,25){\makebox[2mm][c]{$0$}} \put(28,25){\makebox[2mm][c]{$0$}} \put(31,25){\makebox[2mm][c]{$1$}} \put(34,25){\makebox[2mm][c]{$1$}}
\put(29.5,30){\makebox[2mm][c]{$\vdots$}}
\put(-1,-1){ \includegraphics[width=20mm, keepaspectratio=true]{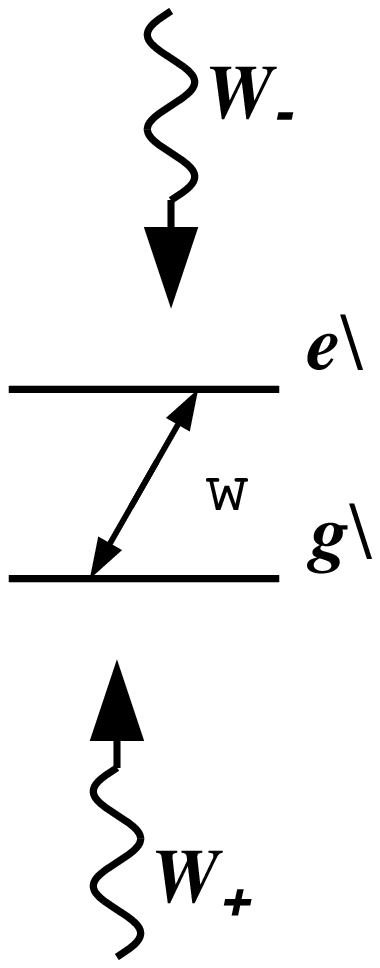} }
\put(61,-1){ \includegraphics[width=25mm, keepaspectratio=true]{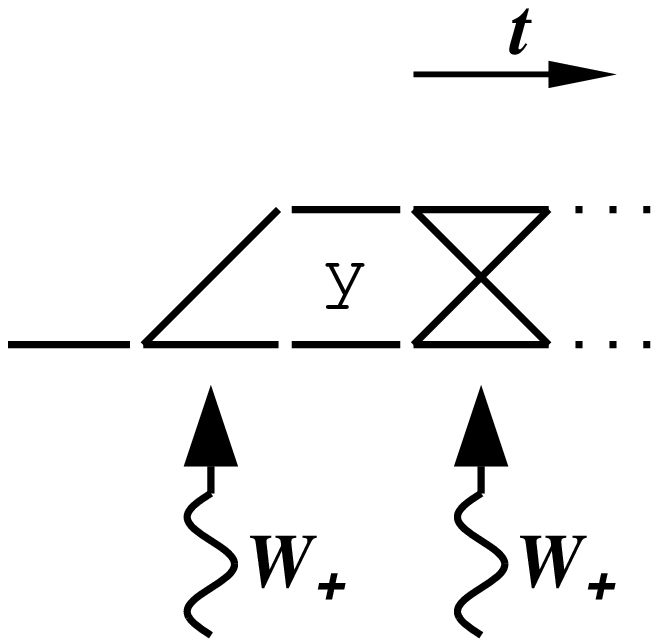} }
\put(0,0){ \parbox[b]{5mm}{(a)} }
\put(22,0){ \parbox[b]{5mm}{(b)} }
\put(61,0){ \parbox[b]{5mm}{(c)} }
\end{picture}
\caption{(a) Fractional $\pi$ pulses tuned to the two-level atom couple adjacent
momentum states, which (b) we label in units of $\hbar k$. This number, in
binary representation, gives the qubits of the momentum state quantum
computer. (c) Two $\pi/2$ pulses may act as the beamsplitters of an atomic
interferometer; the relative phase between the two paths determines whether
the pulses add or subtract, and hence whether or not the electronic state is
inverted.}
\label{Schematic}
\end{figure}

Our scheme is based upon the motion in one dimension (henceforth taken to be
vertical) of a sample of
two-level atoms, such as an atomic beam interacting with transversely
propagating laser beams, as shown in Fig.~\ref{Schematic}. We shall refer to
four coherent operations:

\begin{tabular}{cl}
\hspace{4mm} $W_{+}(\alpha,\phi)$ \hspace{4mm} & a short upward laser pulse \\
$W_{-}(\alpha,\phi)$ & a short downward laser pulse \\
$F(\omega t)$         & free evolution (electronic energy) \\
$G(t/\tau)$            & free evolution (kinetic energy),
\end{tabular}

\noindent where $\tau = 2m / (\hbar k^{2})$. Here, the short (and therefore
spectrally broad) laser pulses couple the upper and lower atomic levels,
between which population is transferred through Rabi cycling for the duration
of the pulse. Conventionally, we describe the overall effect of the pulse
through the phase $2\alpha$ of the Rabi cycle incurred, the population being
inverted when $2\alpha=\pi$ (the so-called `$\pi$ pulse'), and restored when
$2\alpha=2\pi$. Other fractions of a $\pi$ pulse will convert an initially pure
state into a superposition. The phases $\phi$ are determined by the relative
optical phases of the laser pulses. The free evolution operations
$F(\omega t)$ and $G(t/\tau)$ correspond simply to the components of the
time-dependent wavefunction phase $\exp(-{\mathrm i}Et/\hbar)$ that
correspond to the electronic energy and vertical momentum component
respectively. Weitz and H\"{a}nsch~\cite{Weitz00} have shown how the
electronic and kinetic energy contributions to the free phase evolution may
be separated by appropriate insertion of pairs of $\pi$-pulses that invert the
atomic population, as we discuss later.

In this one-dimensional geometry, the atom is constrained to a ladder of
momentum states that are spaced at intervals of the photon momentum
$\hbar k = \hbar \omega / c$ ($\omega$ being the frequency of the resonant
transition) and alternate between the ground and excited electronic levels $g$
and $e$. We label these states according to their momentum components, in
units of the photon impulse $\hbar k$. We initially assume, for our analysis
of the momentum state quantum computer, that these momenta take integer
values, but this assumption will be relaxed when we later consider the
consequences for atomic manipulation.

We now convert our representation of the ladder of states to binary, using
the notation $Q_{n} \ldots Q_{2} Q_{1} Q_{0}$, with the least significant bit
on the right. This is the crucial step in our analysis; yet, apart from the
correspondence of $Q_{0}$ to the electronic state of the atom, binary
representation seems at first rather unpromising, for computational notation
usually helps only when the bits themselves can be manipulated. While the
momentum-changing laser pulses here move population by at most one state
at a time, however, appropriate combinations prove to offer exactly such
bit-wise manipulation.

The key, as in the interferometric cooling scheme of Ref.~\cite{Weitz00}, is
the dependence of the phase of free evolution upon the momentum. For two
levels $p+\Delta p/2$ and $p-\Delta p/2$ (in units of $\hbar k$) and electronic
energy difference $E_{21} = \hbar \omega$, the relative phase $\psi$ evolves
according to
\begin{eqnarray}
\psi & = & \frac{E_{21}t}{\hbar} +
           \frac{(\hbar k)^{2} t}{2 m \, \hbar} \left[ \left(p+\frac{\Delta p}{2} \right)^{2} - \left(p-\frac{\Delta p}{2} \right)^{2} \right] \nonumber \\
& = & \omega t +
           \frac{\hbar k^{2} t}{m} p \, \Delta p.
\end{eqnarray}
Any pair of momentum states thus incur a relative phase that evolves
according to their average momentum $p$. Cancellation of the electronic
contribution to the phase, by inserting $\pi$ pulse pairs so that the
states spend equal times in the ground and excited levels~\cite{Weitz00},
merely changes the average $\Delta p$ and hence the rate at which this
proceeds.

We illustrate the capacity for bit-wise manipulation with the example of a
three qubit right rotation,
\[
\{ Q_{2}, Q_{1}, Q_{0} \} \rightarrow \{ Q_{0}, Q_{2}, Q_{1} \}.
\]
In our largely diagrammatic description, which has its origins in
Fig.~\ref{Schematic}(c), pulses or complete pulse sequences coupling
adjacent levels are indicated by
\setlength{\unitlength}{1pt}
\begin{picture}(9,9)
\put(-1,-2){\line(1,0){10}}
\put(-1,-2){\line(1,1){10}}
\put(-1,8){\line(1,0){10}}
\put(-1,8){\line(1,-1){10}}
\end{picture}
when they produce a superposition (e.g., a $\pi/2$ pulse),
\begin{picture}(9,9)
\put(-1,-2){\line(1,1){10}}
\put(-1,8){\line(1,-1){10}}
\end{picture}
when they cause inversion (the $\pi$ pulse) and
\begin{picture}(9,9)
\put(-1,-2){\line(1,0){10}}
\put(-1,8){\line(1,0){10}}
\end{picture}
when $2\alpha = 0, 2\pi$ and so on. For periods of free evolution, we
simply indicate the relative phases introduced between coupled states.

First consider a pair of upward-travelling $\pi/2$ pulses, separated by a
period of free evolution that introduces between coupled states a relative
phase adjusted to give in each case an integer power of
$\exp(-{\mathrm i}\pi/2)$. This forms a simple interferometer which,
depending upon the original state momentum, can invert population,
return it to its original state, or leave initially pure states coupled (see
Fig.~\ref{Schematic}(c)). The sequence repeats every $8\hbar k$.
\begin{equation}
\raisebox{-16mm}{
\setlength{\unitlength}{1mm}
\begin{picture}(60,37)
{\thicklines \multiput(7,0)(0,5){8}{\line(1,0){5}} }
\put(2,0){ \parbox[b]{5mm}{$0$} }
\put(2,5){ \parbox[b]{5mm}{$1$} }
\put(2,10){ \parbox[b]{5mm}{$2$} }
\put(2,15){ \parbox[b]{5mm}{$3$} }
\put(2,20){ \parbox[b]{5mm}{$4$} }
\put(2,25){ \parbox[b]{5mm}{$5$} }
\put(2,30){ \parbox[b]{5mm}{$6$} }
\put(2,35){ \parbox[b]{5mm}{$7$} }
\put(11,0){ \parbox[b]{5mm}{$|g\rangle$} }
\put(11,5){ \parbox[b]{5mm}{$|e\rangle$} }
\put(11,10){ \parbox[b]{5mm}{$|g\rangle$} }
\put(11,15){ \parbox[b]{5mm}{$|e\rangle$} }
\put(11,20){ \parbox[b]{5mm}{$|g\rangle$} }
\put(11,25){ \parbox[b]{5mm}{$|e\rangle$} }
\put(11,30){ \parbox[b]{5mm}{$|g\rangle$} }
\put(11,35){ \parbox[b]{5mm}{$|e\rangle$} }
\put(20,2.5){ \line(0,1){12.5} }
\put(20,20){ \line(0,1){12.5} }
\put(22.5,2.5){ \oval(5,5)[bl] }
\put(22.5,32.5){ \oval(5,5)[tl] }
\put(17.5,15){ \oval(5,5)[tr] }
\put(17.5,20){ \oval(5,5)[br] }
\multiput(25,0)(0,10){4}{ \line(1,0){5} \line(-1,1){5} }
\multiput(25,5)(0,10){4}{ \line(1,0){5} \line(-1,-1){5} }
\put(36.3,2){ \parbox[b]{5mm}{\large $\frac{\pi}{2}$} }
\put(36.5,12){ \parbox[b]{5mm}{\large $\pi$} }
\put(35.5,22){ \parbox[b]{5mm}{\large $\frac{3 \pi}{2}$} }
\put(35.5,32){ \parbox[b]{5mm}{\large $2 \pi$} }
\multiput(45,0)(0,10){4}{ \line(1,0){5} \line(-1,1){5} }
\multiput(45,5)(0,10){4}{ \line(1,0){5} \line(-1,-1){5} }
\put(55,2.5){ \line(0,1){12.5} }
\put(55,20){ \line(0,1){12.5} }
\put(52.5,2.5){ \oval(5,5)[br] }
\put(52.5,32.5){ \oval(5,5)[tr] }
\put(57.5,15){ \oval(5,5)[tl] }
\put(57.5,20){ \oval(5,5)[bl] }
\end{picture}
}
\equiv
\raisebox{-16mm}{
\setlength{\unitlength}{1mm}
\begin{picture}(5,37)
\multiput(0,0)(0,20){2}{ \line(1,0){5} \line(-1,1){5} }
\multiput(0,5)(0,20){2}{ \line(1,0){5} \line(-1,-1){5} }
\multiput(0,10)(0,5){2}{ \line(1,0){5} }
\put(0,30){ \line(1,1){5} }
\put(0,35){ \line(1,-1){5} }
\end{picture}
}
\label{Stage1}
\end{equation}

\begin{figure}
\includegraphics[width=80mm, keepaspectratio=true]{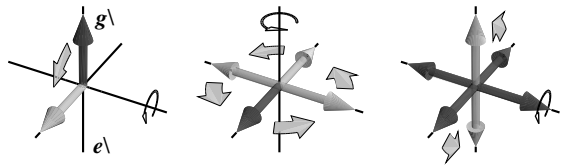}
\caption{Bloch-vector representation of the first stage (Eq.~(\ref{Stage1})) of the
right-rotation. The first pulse rotates the four ground states into
the horizontal plane; free evolution distributes these around the vertical
axis according to their momenta; the second pulse then returns two
to pure states, leaving the others in mixed states. Aside from
phase corrections, the full right-rotation operation takes 18 $\pi/2$
pulses and 26 $\pi$ pulses.}
\label{Vectors}
\end{figure}

This and subsequent operations may be visualized, as in Fig.~\ref{Vectors},
as rotations of Bloch vectors representing the coupled states - a picture
also suitable for NMR computers~\cite{Gershenfeld97,Jones98c}. Pure states
are vertically up ($|g\rangle$) or down ($|e\rangle$), and are coupled by
rotation about a horizontal axis, whose direction depends upon the optical
phase. Free evolution corresponds to rotation about the vertical axis.

Two such sequences may now be combined with a further
momentum-dependent phase between them to form a second interferometer
for the states left in superpositions. The ladder of phases is offset either via
the $F(\omega t)$ operation (such as a period of uncompensated free evolution
that is dominated by the `electronic' phase ), or by appropriate phasing of the
subsequent laser pulses. The result is a conditional state exchange:
\begin{equation}
\raisebox{-16mm}{
\setlength{\unitlength}{1mm}
\begin{picture}(60,37)
{\thicklines \multiput(7,0)(0,5){8}{\line(1,0){5}} }
\put(2,0){ \parbox[b]{5mm}{$0$} }
\put(2,5){ \parbox[b]{5mm}{$1$} }
\put(2,10){ \parbox[b]{5mm}{$2$} }
\put(2,15){ \parbox[b]{5mm}{$3$} }
\put(2,20){ \parbox[b]{5mm}{$4$} }
\put(2,25){ \parbox[b]{5mm}{$5$} }
\put(2,30){ \parbox[b]{5mm}{$6$} }
\put(2,35){ \parbox[b]{5mm}{$7$} }
\put(11,0){ \parbox[b]{5mm}{$|g\rangle$} }
\put(11,5){ \parbox[b]{5mm}{$|e\rangle$} }
\put(11,10){ \parbox[b]{5mm}{$|g\rangle$} }
\put(11,15){ \parbox[b]{5mm}{$|e\rangle$} }
\put(11,20){ \parbox[b]{5mm}{$|g\rangle$} }
\put(11,25){ \parbox[b]{5mm}{$|e\rangle$} }
\put(11,30){ \parbox[b]{5mm}{$|g\rangle$} }
\put(11,35){ \parbox[b]{5mm}{$|e\rangle$} }
\put(20,2.5){ \line(0,1){12.5} }
\put(20,20){ \line(0,1){12.5} }
\put(22.5,2.5){ \oval(5,5)[bl] }
\put(22.5,32.5){ \oval(5,5)[tl] }
\put(17.5,15){ \oval(5,5)[tr] }
\put(17.5,20){ \oval(5,5)[br] }
\multiput(25,0)(0,20){2}{ \line(1,0){5} \line(-1,1){5} }
\multiput(25,5)(0,20){2}{ \line(1,0){5} \line(-1,-1){5} }
\multiput(25,10)(0,5){2}{ \line(1,0){5} }
\put(25,30){ \line(1,1){5} }
\put(25,35){ \line(1,-1){5} }
\put(36.5,2){ \parbox[b]{5mm}{\large $\pi$} }
\put(35.5,12){ \parbox[b]{5mm}{\large $\frac{3 \pi}{2}$} }
\put(35.5,22){ \parbox[b]{5mm}{\large $2\pi$} }
\put(35.53,32){ \parbox[b]{5mm}{\large $\frac{5\pi}{2}$} }
\multiput(45,0)(0,20){2}{ \line(1,0){5} \line(-1,1){5} }
\multiput(45,5)(0,20){2}{ \line(1,0){5} \line(-1,-1){5} }
\multiput(45,10)(0,5){2}{ \line(1,0){5} }
\put(45,30){ \line(1,1){5} }
\put(45,35){ \line(1,-1){5} }
\put(55,2.5){ \line(0,1){12.5} }
\put(55,20){ \line(0,1){12.5} }
\put(52.5,2.5){ \oval(5,5)[br] }
\put(52.5,32.5){ \oval(5,5)[tr] }
\put(57.5,15){ \oval(5,5)[tl] }
\put(57.5,20){ \oval(5,5)[bl] }
\end{picture}
}
\equiv
\raisebox{-16mm}{
\setlength{\unitlength}{1mm}
\begin{picture}(5,37)
\multiput(0,0)(0,5){4}{ \line(1,0){5} }
\put(0,20){ \line(1,1){5} }
\put(0,25){ \line(1,-1){5} }
\multiput(0,30)(0,5){2}{ \line(1,0){5} }
\end{picture}
}
\end{equation}

Precisely which pair of adjacent states is exchanged by this operation
depends upon the directions and relative phases of the $\pi/2$ pulses.
As the penultimate step, we construct the two-qubit exchange operation
{\sf EX(2,1)}:
\begin{equation}
\raisebox{-16mm}{
\setlength{\unitlength}{1mm}
\begin{picture}(60,37)
{\thicklines \multiput(7,0)(0,5){8}{\line(1,0){5}} }
\put(2,0){ \parbox[b]{5mm}{$0$} }
\put(2,5){ \parbox[b]{5mm}{$1$} }
\put(2,10){ \parbox[b]{5mm}{$2$} }
\put(2,15){ \parbox[b]{5mm}{$3$} }
\put(2,20){ \parbox[b]{5mm}{$4$} }
\put(2,25){ \parbox[b]{5mm}{$5$} }
\put(2,30){ \parbox[b]{5mm}{$6$} }
\put(2,35){ \parbox[b]{5mm}{$7$} }
\put(11,0){ \parbox[b]{5mm}{$|g\rangle$} }
\put(11,5){ \parbox[b]{5mm}{$|e\rangle$} }
\put(11,10){ \parbox[b]{5mm}{$|g\rangle$} }
\put(11,15){ \parbox[b]{5mm}{$|e\rangle$} }
\put(11,20){ \parbox[b]{5mm}{$|g\rangle$} }
\put(11,25){ \parbox[b]{5mm}{$|e\rangle$} }
\put(11,30){ \parbox[b]{5mm}{$|g\rangle$} }
\put(11,35){ \parbox[b]{5mm}{$|e\rangle$} }
\put(20,2.5){ \line(0,1){12.5} }
\put(20,20){ \line(0,1){12.5} }
\put(22.5,2.5){ \oval(5,5)[bl] }
\put(22.5,32.5){ \oval(5,5)[tl] }
\put(17.5,15){ \oval(5,5)[tr] }
\put(17.5,20){ \oval(5,5)[br] }
\multiput(25,0)(0,5){3}{ \line(1,0){5} }
\multiput(25,25)(0,5){3}{ \line(1,0){5} }
\put(25,15){ \line(1,1){5} }
\put(25,20){ \line(1,-1){5} }
\multiput(32,0)(0,5){4}{ \line(1,0){5} }
\multiput(32,30)(0,5){2}{ \line(1,0){5} }
\put(32,20){ \line(1,1){5} }
\put(32,25){ \line(1,-1){5} }
\multiput(39,0)(0,5){2}{ \line(1,0){5} }
\multiput(39,20)(0,5){4}{ \line(1,0){5} }
\put(39,10){ \line(1,1){5} }
\put(39,15){ \line(1,-1){5} }
\multiput(46,0)(0,5){3}{ \line(1,0){5} }
\multiput(46,25)(0,5){3}{ \line(1,0){5} }
\put(46,15){ \line(1,1){5} }
\put(46,20){ \line(1,-1){5} }
\put(56,2.5){ \line(0,1){12.5} }
\put(56,20){ \line(0,1){12.5} }
\put(53.5,2.5){ \oval(5,5)[br] }
\put(53.5,32.5){ \oval(5,5)[tr] }
\put(58.5,15){ \oval(5,5)[tl] }
\put(58.5,20){ \oval(5,5)[bl] }
\end{picture}
}
\equiv
\raisebox{-16mm}{
\setlength{\unitlength}{1mm}
\begin{picture}(5,37)
\multiput(0,0)(0,5){2}{ \line(1,0){5} }
\multiput(0,10)(0,5){2}{ \line(1,2){5} }
\multiput(0,20)(0,5){2}{ \line(1,-2){5} }
\multiput(0,30)(0,5){2}{ \line(1,0){5} }
\end{picture}
}
\end{equation}
The right-rotation {\sf RR3} is completed by combining this sequence
with the much simpler exchange {\sf EX(1,0)} of qubits $Q_{1}$ and $Q_{0}$.
This is given in Table~\ref{Opcodes} together with a range of other
operations on the first three qubits from which, in conjunction with the
one-bit operations $W_{+}(\alpha, \phi)$ and $F(\omega t)$,
a complete set may be formed~\cite{Barenco95}. 

\begin{figure}
\includegraphics[width=80mm, keepaspectratio=true]{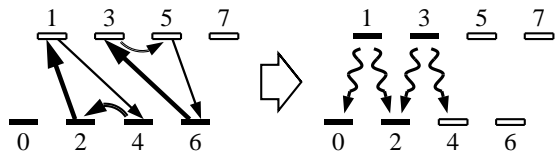}
\caption{Cooling via the right-rotation operation, shown here applied to the
lowest three qubits. The initial distribution across ground states
$\{0, 2, 4, 6 \}$ is transferred  (bold arrows) to the lowest momentum states
$\{0, 1, 2, 3 \}$; subsequent spontaneous emission leaves population in
states $\{0, 2, 4\}$. The width of the momentum distribution may thus be
reduced by a factor of nearly 2 in a single coherent step.}
\label{Rotation}
\end{figure}

When only the ground electronic level is occupied ($Q_{0}=0$), the right-rotation
is indistinguishable from a divide-by-two operation (e.g., $100(=4)\rightarrow 010(=2)$)
and provides a cooling
mechanism. As shown in Fig.~\ref{Rotation}, the four ground states are coherently
mapped onto the four lowest momentum states, two of which subsequently
undergo spontaneous emission, leaving only states $0$, $2$ and $4$ populated.
The coherent process, which pumps heat from kinetic to electronic energy, may
then be repeated, further narrowing the momentum distribution.

\begin{figure}
\includegraphics[width=80mm, keepaspectratio=true]{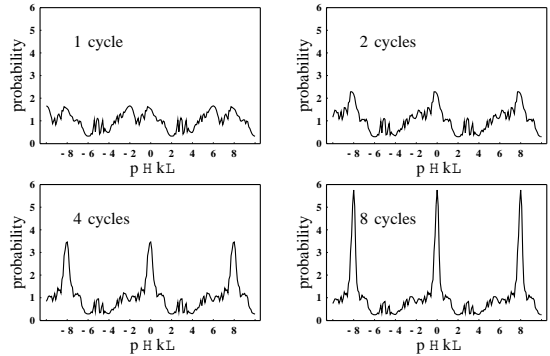}
\caption{Simulated evolution of the momentum state distribution, shown after
1, 2, 4 and 8 cycles of the 3-qubit coherent cooling algorithm. Although the
process applies perfectly only to atoms with even, integral momenta, significant
cooling remains apparent. Spontaneous emission scrambles the exact momenta.
After only a few coherent cycles, the distribution has been narrowed to less than
a single photon impulse. The initial probability density is unity.}
\label{Cooling}
\end{figure}

At this point, we relax our earlier assumption of integer-valued momentum states,
and find that the effect, whilst imperfect for non-integer momenta, nonetheless
remains. The results of our simulations for the effect of this sequence on an
initially flat distribution across fractional momentum states are shown in
Fig.~\ref{Cooling}.

For our simulations, we have used matrix representations of the pulse and evolution
operations. Although the matrices are in principle infinite, all non-zero terms cluster
around the leading diagonal and any element $m_{i,j}$ differs from the diagonally
displaced term $m_{i+2n,j+2n}$ only through its momentum dependence, so we may
summarize the matrices as $4 \times 4$ elements, given below and derived from the
equations of Friedberg and Hartmann~\cite{Friedberg93}. Matrices that we use in
practice must merely be expanded to cover the sequence of interactions and the
momentum range that we wish to describe. The following matrices act on the states
$\{2, 1, 0, -1\}$.

For upward and downward travelling fractional $\pi$ pulses corresponding to Bloch
vector rotation through `Rabi angle' $2\alpha$ and optical phase $\phi$, we have
\begin{equation}
W_{\! +}(\alpha, \phi) = \left( \!\! \begin{array}{cccc}
\cos \alpha & 0 & \hspace{5mm} 0 \hspace{5mm} & \hspace{5mm} 0 \hspace{5mm} \\
0 & \cos \alpha & \!\!\!\! {\mathrm i} e^{{\mathrm i}\phi} \sin \alpha \!\!\!\! & 0 \\
0 & \!\!\!\! {\mathrm i} e^{-{\mathrm i}\phi} \sin \alpha \!\!\!\! & \cos \alpha & 0 \\
\hspace{5mm} 0 \hspace{5mm} & \hspace{5mm} 0 \hspace{5mm} & 0 & \cos \alpha
\end{array} \right)
\end{equation}
and
\begin{equation}
W_{\! -}(\alpha, \phi) = \left( \!\! \begin{array}{cccc}
\cos \alpha & \!\!\!\! {\mathrm i} e^{-{\mathrm i}\phi} \sin \alpha \!\!\!\! & \hspace{5mm} 0 \hspace{5mm} & \hspace{5mm} 0 \hspace{5mm} \\
\!\!\!\! {\mathrm i} e^{{\mathrm i}\phi} \sin \alpha \!\!\!\! & \cos \alpha & 0 & 0 \\
0 & 0 & \cos \alpha & \!\!\!\! {\mathrm i} e^{-{\mathrm i}\phi} \sin \alpha \!\!\!\! \\
\hspace{5mm} 0 \hspace{5mm} & \hspace{5mm} 0 \hspace{5mm} & \!\!\!\! {\mathrm i} e^{{\mathrm i}\phi} \sin \alpha \!\!\!\! & \cos \alpha
\end{array} \right).
\end{equation}
The matrices for free evolution according to the electronic and kinetic energies
are respectively
\begin{equation}
F(\omega t) = \left( \begin{array}{cccc}
1 & 0 & \hspace{4mm} 0 \hspace{4mm} & \hspace{4mm} 0 \hspace{4mm} \\
0 & e^{-{\mathrm i} \omega t} & 0 & 0 \\
0 & 0 & 1 & 0 \\
\hspace{4mm} 0 \hspace{4mm} & \hspace{4mm} 0 \hspace{4mm} & 0 & e^{-{\mathrm i} \omega t}
\end{array} \right)
\end{equation}
and
\begin{equation}
G \left( \frac{t}{\tau} \right) = \left( \begin{array}{cccc}
\!\! e^{- {\mathrm i}(p_{0}+2)^{2} t / \tau} \!\!\!\!\!\!\!\!\!\!\!\! & 0 & \hspace{6mm} 0 \hspace{6mm} & \hspace{6mm} 0 \hspace{6mm} \\
0 & \!\!\!\!\!\!\!\!\!\!\!\! e^{- {\mathrm i}(p_{0}+1)^{2} t / \tau} \!\!\!\!\!\!\!\! & 0 & 0 \\
0 & 0 & \!\!\!\!\!\!\!\! e^{- {\mathrm i}p_{0}^{2} t / \tau} \!\!\!\!\!\!\!\! & 0 \\
\hspace{6mm} 0 \hspace{6mm} & \hspace{6mm} 0 \hspace{6mm} & 0 & \!\!\!\!\!\!\!\! e^{- {\mathrm i}(p_{0}-1)^{2} t / \tau} \!\!
\end{array} \right)
\end{equation}
where $\tau = 2m / (\hbar k^{2})$. Weitz and H\"{a}nsch's sequence for
interferometric cooling thus becomes
\begin{eqnarray*}
W_{\!+} \!\! \left( \! \frac{\pi}{4} \! \right) \! \cdot
G    \!  \left( \frac{2T\!\! -\! T'}{\tau} \right) \! \cdot
F     \!  \left( \omega \left(2T\!\! -\! T' \right) \right) \cdot
W_{\!-} \!\!  \left( \! \frac{\pi}{2} \! \right) \! \cdot
G     \!  \left( \frac{2T}{\tau} \right) \! \cdot \\
F     \!  \left( 2 \omega T \right) \cdot
W_{\!-} \!\!  \left( \! \frac{\pi}{2} \! \right) \! \cdot
G     \!  \left( \frac{T'}{\tau} \right) \! \cdot
F     \!  \left( \omega T' \right) \cdot
W_{\!+} \!\!  \left( \! \frac{\pi}{4} \! \right)
\end{eqnarray*}

No attempt has yet been made to optimize the sequences given here: the
phase corrections often serve only to demonstrate the exact equivalence to
a quantum computer, and reductions in the complexity, duration or
momentum sensitivity of each operation should be possible. Nor have we
examined superpositions involving more than two states~\cite{Hinderthuer99}
or interactions at more than one wavelength~\cite{Soeding97}. Instead of the
fractional Rabi coupling of electronic transitions assumed for simplicity, our
scheme could be more robustly implemented using Raman
transitions~\cite{Weiss93} for adiabatic passage~\cite{Weitz94,Featonby98}
between Zeeman or hyperfine levels, possible even with modulated c.w.
lasers~\cite{Cashen02}. The scheme, which we think of as a form of
`algorithmic cooling'~\cite{Boykin02} in its broadest sense, could in principle be extended to
three dimensions. Owing to the non-resonant nature of the pulsed
interactions, it would  also be suitable for molecules, for which the large
impulse per coherent cycle would be a particular advantage.

\newpage
\begin{table*}
\caption{Basic operations of the momentum state quantum computer. No attempt has
been made to optimize the pulse sequences, which run from right to left, and some
uncorrected phases remain in the operations marked with an asterisk. For the operation
 $G$, $p_{0}$ is taken to be zero (mod 8).}
\label{Opcodes}
\begin{ruledtabular}
\begin{tabular}{llll}
level & name & description & sequence \\ \hline
basic & $G(t / \tau)$ & & $
    W_{\! -} \! \left( \pi, 0 \right) \cdot FG(t / 4 \tau) \cdot
    W_{\! -} \! \left( \pi, 0 \right) \cdot FG(t / 4 \tau) \cdot $ \\
&&& \hspace{3mm} $
    W_{\! +} \! \left( \pi, 0 \right) \cdot FG(t / 4 \tau) \cdot
    W_{\! +} \! \left( \pi, 0 \right) \cdot FG(t / 4 \tau)$ \\ \hline
1 qubit & {\sf NOT(0)} & $Q_{0} \rightarrow \overline{Q_{0}}$ & $
    F \! \left( \frac{\pi}{2} \right) \cdot
    W_{\! +} \! \left( \frac{\pi}{2}, 0 \right) \cdot
    F \! \left( \frac{\pi}{2} \right)$ \\
& {\sf CP1(0)} & if state=$0$, invert phase &  $
    F \! \left( \pi \right) \cdot
    W_{\! +} \! \left( \pi, 0 \right)$ \\
& {\sf HAD(0)} & Walsh-Hadamard on $Q_{0}$ &  $
    W_{\! +} \! \left( \frac{\pi}{4}, \frac{\pi}{2} \right) \cdot
    F \! \left( \pi \right) \cdot
    W_{\! +} \! \left( \pi, 0 \right)$ \\ \hline
2 qubit & {\sf EX(1,0)} & $\{Q_{1}$, $Q_{0}\} \rightarrow \{Q_{0}$, $Q_{1}\}$ & $
    F \! \left( \frac{\pi}{2} \right) \cdot
    W_{\! -} \! \left( \frac{\pi}{4}, \pi \right) \cdot
    G \! \left( \frac{\pi}{4} \right) \cdot
    W_{\! -} \! \left( \frac{\pi}{4}, \frac{\pi}{4} \right) \cdot
    F \! \left( \frac{5 \pi}{4} \right)$ \\
& {\sf CNOT(1,0)} & $\{Q_{1},Q_{0}\}\rightarrow \{Q_{1},Q_{1} \oplus Q_{0}\}$ & $
    F \! \left( \frac{\pi}{2} \right) \cdot
    W_{\! +} \! \left( \frac{\pi}{4}, \pi \right) \cdot
    G \! \left( \frac{\pi}{4} \right) \cdot
    W_{\! +} \! \left( \frac{\pi}{4}, \frac{\pi}{4} \right) \cdot
    F \! \left( \frac{5 \pi}{4} \right)$ \\
& $\overline{\mbox{\sf CNOT}}\mbox{\sf (1,0)}$ & $\{Q_{1},Q_{0}\}\rightarrow \{Q_{1},\overline{Q_{1} \oplus Q_{0}}\}$ & $
    W_{\! +} \! \left( \pi, 0 \right) \cdot
    F \! \left( \frac{3 \pi}{2} \right) \cdot
    W_{\! +} \! \left( \frac{\pi}{4}, 0 \right) \cdot
    G \! \left( \frac{\pi}{4} \right) \cdot
    W_{\! +} \! \left( \frac{\pi}{4}, \frac{\pi}{4} \right) \cdot
    F \! \left( \frac{5 \pi}{4} \right)$ \\
& {\sf CP2(0)} & if state=$0$, invert phase & $
    F \! \left( \frac{3 \pi}{4} \right) \cdot
    G \! \left( \frac{\pi}{4} \right) \cdot
    W_{\! +} \! \left( \pi, 0 \right)$ \\
& {\sf HAD(1,0)} & Walsh-Hadamard on $Q_{1}$, $Q_{0}$ &
    {\sf EX(1,0)} $\cdot$
    {\sf HAD(0)} $\cdot$
    {\sf EX(1,0)} $\cdot$
    {\sf HAD(0)} \\ \hline
3 qubit & {\sf SW3(2,3)$^{*}$} & swap states 2, 3 & $
     W_{\! +} \! \left( \frac{\pi}{4}, 0 \right) \cdot
     G \! \left( \frac{\pi}{8} \right) \cdot
     W_{\! +} \!  \left( \frac{\pi}{4}, \frac{9 \pi}{8} \right) \cdot
     F \! \left( \frac{5 \pi}{4} \right) \cdot
     G \! \left( \frac{\pi}{8} \right) \cdot $ \\
&&& \hspace{3mm} $
     W_{\! +} \! \left( \frac{\pi}{4}, 0 \right) \cdot
     G \! \left( \frac{\pi}{8} \right) \cdot
     W_{\! +} \! \left( \frac{\pi}{4}, \frac{9 \pi}{8} \right) \cdot
     F \! \left( \frac{13 \pi}{8} \right) \cdot
     G \! \left( \frac{\pi}{4} \right)$ \\
& {\sf SW3(3,4)$^{*}$} & swap states 3, 4 & $
    F \! \left( \pi \right) \cdot
    W_{\! -} \! \left( \frac{\pi}{4}, 0 \right) \cdot
    G \! \left( \frac{\pi}{8} \right) \cdot
    W_{\! -} \! \left( \frac{\pi}{4}, \frac{5 \pi}{8} \right) \cdot
    F \! \left( \frac{5 \pi}{4} \right) \cdot
    G \! \left( \frac{\pi}{8} \right) \cdot $ \\
&&& \hspace{3mm} $
    W_{\! -} \! \left( \frac{\pi}{4}, 0 \right) \cdot
    G \! \left( \frac{\pi}{8} \right) \cdot
    W_{\! -} \! \left( \frac{\pi}{4}, \frac{5 \pi}{8} \right) \cdot
    F \! \left( \frac{13 \pi}{8} \right) \cdot
    G \! \left( \frac{\pi}{4} \right)
   $ \\
& {\sf SW3(4,5)$^{*}$} & swap states 4, 5 & $
    W_{\! +} \! \left( \frac{\pi}{4}, 0 \right) \cdot
    G \! \left( \frac{\pi}{8} \right) \cdot
    W_{\! +} \! \left( \frac{\pi}{4}, \frac{5 \pi}{8} \right) \cdot
    F \! \left( \frac{5 \pi}{4} \right) \cdot
    G \! \left( \frac{\pi}{8} \right) \cdot $ \\
&&& \hspace{3mm} $
    W_{\! +} \! \left( \frac{\pi}{4}, 0 \right) \cdot
    G \! \left( \frac{\pi}{8} \right) \cdot
    W_{\! +} \! \left( \frac{\pi}{4}, \frac{5 \pi}{8} \right) \cdot
    F \! \left( \frac{\pi}{8} \right) \cdot
    G \! \left( \frac{\pi}{4} \right)
    $ \\
& {\sf EX(2,1)} & $\{Q_{2}, Q_{1}\} \rightarrow \{Q_{1}, Q_{2}\}$ & $
    W_{\! +} \! \left( \pi, 0 \right) \cdot
    \overline{\mbox{\sf CNOT}}\mbox{\sf (1,0)} \cdot
    \mbox{\sf EX(1,0)} \cdot
    G \! \left( \frac{3 \pi}{8} \right) \cdot
    F \! \left( \frac{13 \pi}{8} \right) \cdot
    \mbox{\sf EX(1,0)} \cdot
    \overline{\mbox{\sf CNOT}}\mbox{\sf (1,0)} \cdot $ \\
&&& \hspace{3mm} $
    \mbox{\sf SW3(3,4)} \cdot
    \mbox{\sf NOT(0)} \cdot
    F \! \left( \pi \right) \cdot
    \mbox{\sf NOT(0)} \cdot
    \mbox{\sf SW3(4,5)} \cdot$ \\
&&& \hspace{3mm} $
    \mbox{\sf NOT(0)} \cdot
    F \! \left( \pi \right) \cdot
    \mbox{\sf NOT(0)} \cdot 
    \mbox{\sf SW3(2,3)} \cdot
    \mbox{\sf SW3(3,4)} \cdot
    G \! \left( \frac{3 \pi}{8} \right) \cdot
    F \! \left( \frac{5 \pi}{8} \right) $ \\
& {\sf RR3} & $\{Q_{2}, Q_{1}, Q_{0}\} \rightarrow \{Q_{0}, Q_{2}, Q_{1}\}$ & {\sf EX(2,1)} $\cdot$ {\sf EX(1,0)} \\
& {\sf RL3} & $\{Q_{2}, Q_{1}, Q_{0}\} \rightarrow \{Q_{1}, Q_{0}, Q_{2}\}$ & ${\sf RR3} \cdot {\sf RR3}$ \\
& {\sf CP3(0)} & if state=$0$, invert phase & $
    {\sf NOT(0)} \cdot
    {\sf RL3} \cdot
    {\sf NOT(0)} \cdot
    {\sf RL3} \cdot
    F \! \left( \frac{5 \pi}{8} \right) \cdot
    G \! \left( \frac{3 \pi}{8} \right) \cdot
    {\sf RR3} \cdot
    {\sf SW3(4,5)} \cdot
    F \! \left( \frac{3 \pi}{2} \right) \cdot $ \\
&&& \hspace{3mm} $
    {\sf SW3(4,5)} \cdot
    F \! \left( \frac{\pi}{2} \right) \cdot
    {\sf NOT(0)} \cdot
    {\sf RR3} \cdot
    {\sf NOT(0)} \cdot
    W_{\! +} \! \left( \pi, 0 \right) $
\end{tabular}
\end{ruledtabular}
\end{table*}

\newcommand{\noopsort}[1]{} \newcommand{\printfirst}[2]{#1}
  \newcommand{\singleletter}[1]{#1} \newcommand{\switchargs}[2]{#2#1}

\end{document}